\documentclass[aps,prb,twocolumn,amsmath,amssymb,superscriptaddress]{revtex4-1}

\usepackage{graphicx}
\usepackage{bm}
\usepackage{mathptmx}
\usepackage[english]{babel}
\usepackage{indentfirst}
\usepackage{color}
\usepackage{listings}
\usepackage{multibib}
\usepackage{epstopdf}

\begin{document}

\title{Impact of high-frequency pumping on anomalous finite-size effects in three-dimensional topological insulators}

\author{Anastasiia A. Pervishko}
\affiliation{ITMO University, Saint Petersburg 197101, Russia}

\author{Dmitry Yudin}
\affiliation{ITMO University, Saint Petersburg 197101, Russia}

\author{Ivan A. Shelykh}
\affiliation{ITMO University, Saint Petersburg 197101, Russia}
\affiliation{Science Institute, University of Iceland, Dunhagi 3, Reykjavik IS-107, Iceland}

\begin{abstract}
Lowering of the thickness of a thin-film three-dimensional topological insulator down to a few nanometers results in the gap opening in the spectrum of topologically protected two-dimensional surface states. This phenomenon, which is referred to as the anomalous finite-size effect, originates from hybridization between the states propagating along the opposite boundaries. In this work, we consider a bismuth-based topological insulator and show how the coupling to an intense high-frequency linearly polarized pumping can further be used to manipulate the value of a gap. We address this effect within recently proposed Brillouin-Wigner perturbation theory that allows us to map a time-dependent problem into a stationary one. Our analysis reveals that both the gap and the components of the group velocity of the surface states can be tuned in a controllable fashion by adjusting the intensity of the driving field within an experimentally accessible range and demonstrate the effect of light-induced band inversion in the spectrum of the surface states for high enough values of the pump.
\end{abstract}

\maketitle

\section{Introduction} 

The use of topological ideas has become increasingly ubiquitous in modern condensed-matter physics \cite{Thouless1982,Pankratov1987,Wen1995,Hasan2010,Qi2011} since the pioneering work on the quantum Hall effect \cite{Klitzing1980}. Particularly, the theory of phase transitions was dramatically revolutionized with the advent of the concept of topological order: the formation of phases of matter is no longer associated with the emergence of spontaneous symmetry breaking and corresponding order parameter, but with the emergence of a topological invariant in Hilbert space that is specified uniquely by the properties of the occupied Bloch states \cite{Senthil2015,Bansil2016}. Furthermore, the subsequent observation of quantum spin Hall effect in certain semiconducting structures with strong spin-orbit coupling \cite{Kane2005,Bernevig2006,Konig2007,Hsieh2008} has intensified the search for different topological phases of matter. Among them is the recently predicted \cite{Kane2005a,Fu2007,Fu2007a,Moore2007,Roy2009} and experimentally realized \cite{Xia2009,Zhang2009,Chen2009} three-dimensional topological insulator (TI): a system which behaves like an insulator in the three-dimensional bulk but supports gapless conducting electronic modes at the two-dimensional boundaries protected by time-reversal symmetry. The conduction is attributed to the presence of extended edge or surface states which possess a remarkable property of topological protection. It is predicted that the electrons traveling along the surface are immune against the backscattering and preserve their quantum phase coherence over a long distance despite the presence of impurities, interactions, and external fields \cite{Hasan2010,Qi2011,Gehring2015}.
Despite enormous progress in the study of TIs they still remain an extremely active area of research due to unique electronic and optical properties allowing potential applications in the emerging domain of quantum technologies \cite{Vobornik2011,Pesin2012}.

Typically, Bi$_2$Se$_3$, representing a system of stacked quintuple layers with rather weak coupling between them, is considered a paradigmatic strong three-dimensional TI. The formation of a single Dirac cone in this compound has been clearly justified with angle-resolved photoemission spectroscopy (ARPES) \cite{Roushan2009} and scanning tunneling microscopy (STM) \cite{Alpichshev2010} measurements. Being deposited in the form of thin films the spectrum of surface states of this material was shown to be gapful, if the thickness of a layer is below 10 nm due to certain interference between top and bottom states \cite{Linder2009,Zhang2010,Lu2010,Liu2010}. Further increase of the layer thickness results in the gap collapse. 

At the same time, using the coupling to an external electromagnetic field opens up interesting opportunities to study exotic states of hybrid light-matter nature \cite{Oka2009,Inoue2010,Karch2010,Morell2012,Rudner2013,Grushin2014,Kundu2014,Titum2015,Dahlhaus2015,Klinovaja2016,Sato2016}. For highly intense lasers light-matter coupling is sufficiently strong to cause the breakdown of standard perturbative expansion. In such cases, the most suitable approach is based on Floquet-Magnus-type expansion \cite{Casas2001,Blanes2009,Mananga2011} allowing us to map the time-dependent problem into an effective stationary one. Topological classification has been further generalized to the systems driven out of equilibrium with time-periodic pumping \cite{Kitagawa2010}, leading, in particular, to the concept of Floquet TI \cite{Lindner2011,Dora2012,Rechtsman2013,Wang2013,Sentef2015}. In the latter case, a system exhibits a single pair of helical edge modes in the gap of its quasienergy spectrum due to the light-induced band inversion. It turns out that even a conventional band insulator can be driven to a topologically nontrivial phase with on-resonant field which gives rise to the Floquet band reshuffling \cite{Lindner2011}. 

Furthermore, the idea of utilizing off-resonant pumping to drive low-dimensional electronic systems to nontrivial topological phases has also been pushed forward \cite{Kitagawa2011,Tahir2016,Kuwahara2016,Mikami2016,Hasan2017}. For high-frequency electromagnetic fields the real processes of photon absorption or emission cannot occur because of the constraints imposed by the energy conservation. However, off-resonant light can affect the electron system via virtual processes, leading thus to the pronounced band-structure renormalization \cite{Wang2014,Morina2015,Pervishko2015,Kristinsson2016,Yudin2016,Yudin2016a,Kibis2017}. Meanwhile, almost no attention was paid to the idea of using high-frequency electromagnetic radiation to the finite-size systems despite its potential impact on the realization of all-optical control in topological systems. In this paper we show that light-matter coupling leads to substantial renormalization of the bare Hamiltonian of a three-dimensional TI. We develop a theoretical description of a thin slab of Bi$_2$Se$_3$ driven by intense high-frequency pumping within  Brillouin-Wigner perturbation-theory approach \cite{Mikami2016} to show that the anomalous finite-size effects in the spectrum of surface states are sensitive to the intensity of the driving. The rest of the paper is organized as follows: In Sec. \ref{sec:hamiltonian} we derive an effective time-independent Hamiltonian, study dispersion relation of corresponding surface states for a semi-infinite sample in Sec. \ref{subsec:infinite} and a finite slab of TI in Sec. \ref{subsec:finite} and demonstrate that the values of the gap and the components of the group velocity of the surface states can be manipulated by purely optical means. We summarize our findings in Sec. \ref{sec:conclusions}.

\begin{figure}[h]
\begin{center}
\includegraphics[scale=1.2]{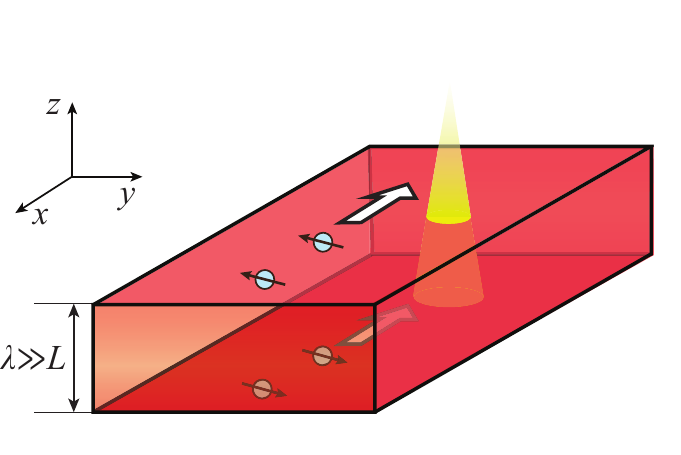}
\caption{\label{model}(Color online) Geometry of the system discussed in the text: a slab of three-dimensional TI of the thickness $L$ is pumped with an intense linearly polarized field propagating normally to its surface. Throughout the calculations we assume $L\ll\lambda$, the wavelength of the driving, which validates neglecting coordinate dependence of the field.}
\end{center}
\end{figure}

\section{Effective time-independent Hamiltonian}\label{sec:hamiltonian}

We start our analysis with the four-band model of a three-dimensional TI in the vicinity of the $\Gamma$ point. The corresponding Hamiltonian can be derived within $\mathbf{k}\cdot\mathbf{p}$ theory with a proper account of corresponding crystalline symmetries,
\begin{equation}\label{ham}
H=\mathcal{E}_\mathbf{k}+\left(\begin{array}{cc}
\mathcal{M}_\mathbf{k}\hat{\tau}_z+A_1k_z\hat{\tau}_x & A_2k_-\hat{\tau}_x \\
A_2k_+\hat{\tau}_x & \mathcal{M}_\mathbf{k}\hat{\tau}_z-A_1k_z\hat{\tau}_x
\end{array}\right),
\end{equation}
where the three-dimensional wave vector $\mathbf{k}=(k_x,k_y,k_z)$ specifies the electron states, $\mathcal{E}_\mathbf{k}=C+D_1k_z^2+D_2k_\perp^2$ (without loss of generality, in the following we put $D_2=0$), $\mathcal{M}_\mathbf{k}=B_0-B_1k_z^2-B_2k_\perp^2$, with $k_\pm=k_x\pm ik_y$ and $k_\perp^2=k_x^2+k_y^2$, while the parameters $A_1$, $A_2$, $B_0$, $B_1$, $B_2$, $D_1$, and $C$ have to be fit to the results of {\it ab initio} simulations \cite{Zhang2009}. The Hamiltonian (\ref{ham}) is written in the basis of hybridized $p$-orbital states of Bi ($P1$) and Se ($P2$) with a given parity ($\pm$) and spin projection $\left\lbrace\vert P1_z^+,\uparrow\rangle,\vert P2_z^-,\uparrow\rangle,\vert P1_z^+,\downarrow\rangle,\vert P2_z^-,\downarrow\rangle\right\rbrace$. The former expansion is legitimate in a narrow region around the Fermi level. In the expression (\ref{ham}) we defined a set of Pauli matrices $\hat{\tau}_x$ and $\hat{\tau}_z$. 

We work in the geometry schematically shown in Fig.~\ref{model}. Assume the TI is pumped with a time-periodic electromagnetic field $\mathbf{E}=\mathbf{E}_0\sin\omega t$ propagating along the $z$ axis normal to the surface of TI. With extremely high accuracy for an ultrathin system $L\ll\lambda$, where $L$ is the thickness of the layer of TI and $\lambda$ corresponds to the wavelength of the driving, one can neglect the variation of the field in the $z$ direction inside the bulk  of TI. The driving field is supposed to be linearly polarized along $y$ axis, $\mathbf{E}_0=E_0\hat{\bm{y}}$, with the amplitude $E_0$ and the frequency $\omega$. Time dependence is induced to the Hamiltonian (\ref{ham}) via the minimal coupling to the field $\mathbf{k}\rightarrow\mathbf{k}-\gamma\hat{\bm{y}}\cos\omega t$, where $\gamma=eE_0/(\hbar\omega)$. In the high-frequency regime no real absorption or emission of light quanta can happen --- electrons cannot follow rapid oscillations of the driving field. Meantime, one may write down an effective time-independent Hamiltonian using the high-frequency expansion in the form of the Brillouin-Wigner perturbative approach that has been recently suggested for this class of problems \cite{Mikami2016}.

In order to obtain the renormalized parameters of the Hamiltonian (\ref{ham}) we proceed by estimating the terms proportional to $1/\omega^2$ (see Appendix \ref{appendix:expansion} for computational details). Similar to \cite{Hasan2017} for a given field polarization the contribution due to $1/\omega$ vanishes, and we arrive at
\begin{equation}\label{renham}
H_\mathrm{eff}=\mathcal{E}_\mathbf{k}+\left(\begin{array}{cc}
\tilde{\mathcal{M}}_\mathbf{k}\hat{\tau}_z+\tilde{A}_1k_z\hat{\tau}_x & (\tilde{A}_{2x}k_x-i\tilde{A}_{2y}k_y)\hat{\tau}_x \\
(\tilde{A}_{2x}k_x+i\tilde{A}_{2y}k_y)\hat{\tau}_x & \tilde{\mathcal{M}}_\mathbf{k}\hat{\tau}_z-\tilde{A}_1k_z\hat{\tau}_x
\end{array}\right),
\end{equation}
with $\tilde{A}_1=A_1\left(1-a-b\right)$, $\tilde{A}_{2x}=A_2\left(1-a-b\right)$, $\tilde{A}_{2y}=A_2\left(1-2aB_0B_2/A_2^2+7b\right)$, and $\tilde{\mathcal{M}}_\mathbf{k}=\tilde{M}-\tilde{B}_1k_z^2-\tilde{B}_2k_\perp^2-2aB_2k_y^2$, on the condition that $a=\left[A_2\gamma/(\hbar\omega)\right]^2$, $b=\left[B_2\gamma^2/(4\hbar\omega)\right]^2$, $\tilde{B}_i=(1-a)B_i$, and $\tilde{M}=\tilde{B}_0-(2-a)B_2\gamma^2/4$. Indeed, our calculations unambiguously determine renormalization of the parameters of the Hamiltonian. The comparison with the noninteracting case (\ref{ham}) shows that the effect of renormalization is substantial. Noteworthy, we consider the effect of high-frequency driving on low-energy bands exclusively and suppose that optical processes that somehow involve high-energy bands have no pronounced consequences on low-energy states.

\section{Band structure of surface states}\label{sec:surface} 

In this section we discuss the dispersion relation of surface states of a three-dimensional TI described by the Hamiltonian (\ref{renham}). We consider two different geometries: semi-infinite TI and finite slab of thickness $L$. It turns out that the former case allows analytical treatment, whereas for the latter we provide the results of the numerical simulations. 

\subsection{Semi-infinite geometry}\label{subsec:infinite}

Remarkably, for a certain geometry we can make use of the Hamiltonian (\ref{renham}) to obtain analytically the spectrum of the surface states. We consider a three-dimensional TI, occupying the semispace $z<0$ and contacting with a standard band insulator ($z>0$), so that at the boundary $z=0$ surface states are present, provided that the wave function $\psi(x,y,z=0)=0$. In the vicinity of the $\Gamma$ point ($\mathbf{k}=0$) the Hamiltonian allows a pair of doubly degenerate surface states. To gain further analytical insight we put $C=D_1=0$. Thanks to the translational symmetry along $x$ and $y$ directions components of quasimomentum $k_x$ and $k_y$ are the good quantum numbers, whereas $k_z\rightarrow -i\partial_z$ has to be replaced, leading thus to the Schr\"odinger equation $\tilde{H}(-i\partial_z)\psi(z)=\varepsilon\psi(z)$ at the $\Gamma$ point. Two independent doubly degenerate solutions of this equation are $\psi_i(z)=(e^{-\kappa_+z}-e^{-\kappa_-z})|\psi_i\rangle$ ($i=1,2$) with
\begin{equation}
|\psi_1\rangle=\frac{1}{\sqrt{2}}\left(\begin{array}{c}
1 \\ i \\ 0 \\ 0
\end{array}\right), \quad 
|\psi_2\rangle=\frac{1}{\sqrt{2}}\left(\begin{array}{c}
0 \\ 0 \\ -1 \\ i
\end{array}\right),
\end{equation}
where we defined
\begin{equation}
\kappa_\pm=\left[f_1-f_2\pm\sqrt{f_1(f_1-2f_2)}\right]^{1/2},
\end{equation}
provided that the values of $f_1=\tilde{A}_1^2/(2\tilde{B}_1^2)$ and $f_2=\tilde{M}/\tilde{B}_1$ are determined by the parameters of the renormalized Hamiltonian (\ref{renham}). If we project the effective time-independent Hamiltonian (\ref{renham}) to the subspace spanned by the surface states $\{\psi_1(z),\psi_2(z)\}$, and keep the terms linear in $k_x$ and $k_y$ we derive the standard Hamiltonian of surface states with the velocities renormalized by the electromagnetic field,
\begin{equation}\label{surham}
H=\hbar\left(v_yk_y\hat{\sigma}_x-v_xk_x\hat{\sigma}_y\right),
\end{equation}
where $v_x=\tilde{A}_{2x}/\hbar$ and $v_y=\tilde{A}_{2y}/\hbar$. The set of Pauli matrices $\hat{\sigma}_x$ and $\hat{\sigma}_y$ act in spin space. The expression (\ref{surham}) reproduces the celebrated result $v_0=v_x=v_y=A_2/\hbar$ in the absence of external pumping. Meanwhile, the presence of the driving makes the motion of electrons along $x$ and $y$ directions inequivalent. To estimate how pronounced the predicted effect is we make use of the following parameters: $\hbar\omega=1$ eV and $eE_0=0.1$ eV\,\AA$^{-1}$, which can be achieved in up-to-date experimental facilities. In particular, for Bi$_2$Se$_3$ this results in the velocity $v_x=0.94\,v_0$ and $v_y=1.06\,v_0$ being anisotropic and slightly bigger in the direction of field polarization.

\begin{figure}[h]
\includegraphics[scale=0.75]{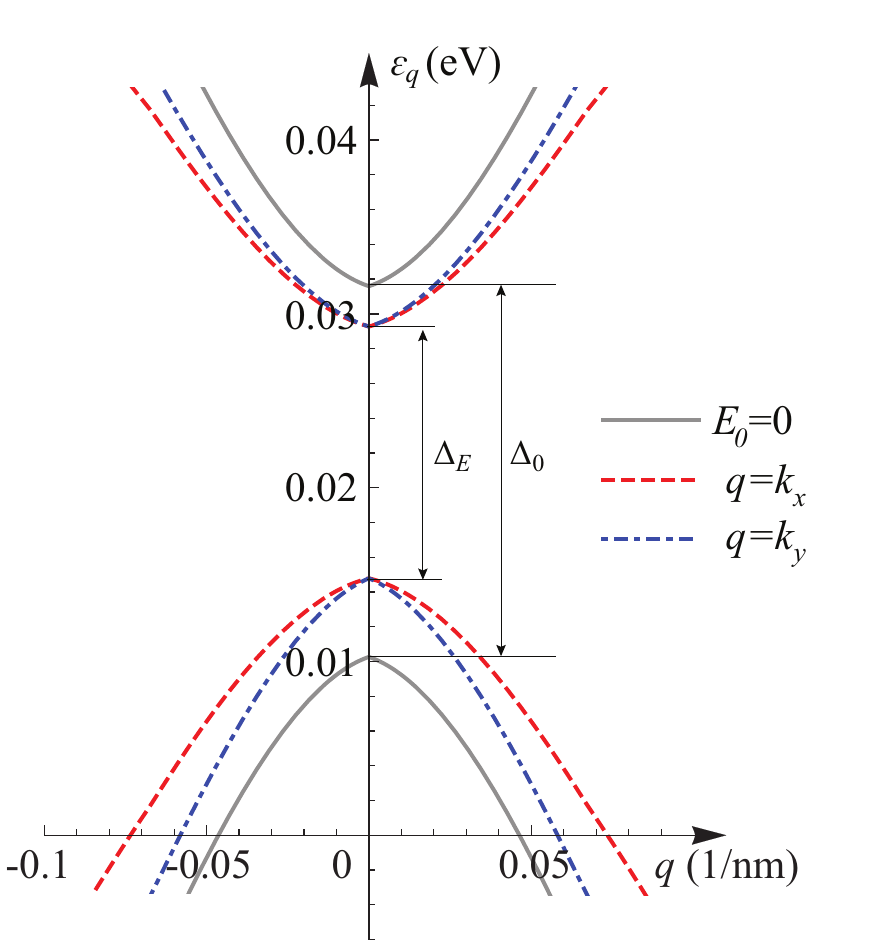}
\caption{\label{dispersion}(Color online) Dispersion relation of topologically protected surface states of a three-dimensional TI of thickness $L=4$ nm in the absence of the field (marked with black solid lines) and for the field with the amplitude $E_0=1.7\times10^8$ V/m and frequency $\omega=600$ THz (marked with red dashed line at $k_y=0$ and blue dash-dotted line at $k_x=0$). It is clear that the value of a gap in the spectrum can be tuned by changing the parameters of the field. The value of the field-renormalized gap $\Delta_E$ is smaller than that of the bare one, $\Delta_0$.}
\end{figure}
  
Instructively, if one starts directly with the Hamiltonian of the surface states $\hbar v_0\left(\hat{\sigma}\times\mathbf{k}\right)_z$ and develops high-frequency expansion, one can clearly see that only $v_x$ is renormalized \cite{Yudin2016}. The latter results from the fact that the only relevant parameter in this model is the Fermi velocity at the Dirac point which inherits the microscopic parameter $A_2$ exclusively. The results of our derivation in this case suggest that $v_x=v_0\left[1-(v_0\gamma/\omega)^2\right]$ and $v_y=v_0$ which is in line with Ref. \cite{Yudin2016}. 

\subsection{Thin-film topological insulator}\label{subsec:finite}

Consider an ultrathin TI of the thickness $L$ in $z$ direction and infinite in $x$ and $y$ directions as schematically shown in Fig.~\ref{model}. To identify the spectrum of surface states we solve the Schr\"odinger equation with the renormalized Hamiltonian upon imposing hard-wall, or surface, boundary conditions \cite{Zhou2008,Linder2009,Lu2010,Liu2010,Silvestrov2012}, according to which the wave function $\psi(x,y;z=0,L)=0$ at the surfaces where TI terminates. Note that in our analysis we do not consider correcting surface potentials as proposed in Ref. \cite{Zhang2012}. Working out the eigenvalue problem with the Hamiltonian (\ref{renham}) leads to the dispersion relation in the form (see Appendix \ref{appendix:dispersion} for derivation)
\begin{equation}\label{disp1}
\sum\limits_\pm\left[\frac{\tanh(\lambda_\pm L/2)}{\tanh(\lambda_\mp L/2)}-\frac{\lambda_\pm}{\lambda_\mp}\left(1\pm\frac{D_1^2-B_1^2}{D_1^2-\tilde{B}_1^2}\frac{\sqrt{R}}{\tilde{A}_1^2}\right)\right]=0,
\end{equation}
provided that
\begin{equation}\label{disp2}
\lambda_\pm=\left[\frac{F\pm\sqrt{R}}{2(D_1^2-\tilde{B}_1^2)}\right]^{1/2},
\end{equation}
where $F=(l_++l_-)D_1+(l_+-l_-)\tilde{B}_1-\tilde{A}_1^2$, $R=F^2-4(D_1^2-\tilde{B}_1^2)[l_+l_--(\tilde{A}_{2x}k_x)^2-(\tilde{A}_{2y}k_y)^2]$, and $l_\pm=C\pm\tilde{\mathcal{M}}_\mathbf{k}\mp\tilde{B}_2k_\perp^2-\varepsilon$.

Equation (\ref{disp1}) implicitly determines spectrum of the surface states $\varepsilon_\mathbf{k}(L)$ of a thin-film TI depending on the thickness $L$ of a layer. In contrast to the case with no radiation present in the system for which the energy dispersion is purely defined by $k_\perp$, the presence of light makes the system in question effectively anisotropic in ($k_x,k_y$) space as shown in Fig.~\ref{dispersion} (we discuss the details of numerical simulations in Appendix \ref{appendix:dispersion}). When the thickness of a layer decreases down to a few nanometers a gap opens up at the $\Gamma$ point and top and bottom surface states become hybridized. The latter suggests a wide range of potential applications of Bi$_2$Se$_3$ as such a structure preserves its nice topological properties before finite-size effects come into play at $L\leq 10$ nm, which is in marked contrast with its two-dimensional counterparts \cite{Zhou2008}. It was recently shown that the latter stems from the rather short localization length of the topologically protected states which results from a very large charge-excitation gap in the bulk. Noteworthy, in our numerical simulations to model the properties of Bi$_2$Se$_3$ we utilize the following parameters \cite{Zhang2009}: $C=6.8\times10^{-3}$ eV, $B_0=0.28$ eV, $A_1=2.2$ eV \AA, $A_2=4.1$ eV \AA, $B_1=10$ eV \AA$^2$, $B_2=56.6$ eV \AA$^2$, $D_1=1.3$ eV \AA$^2$, and $D_2=19.6$ eV \AA$^2$.

\begin{figure}[htb]
\begin{center}
\includegraphics[scale=0.47]{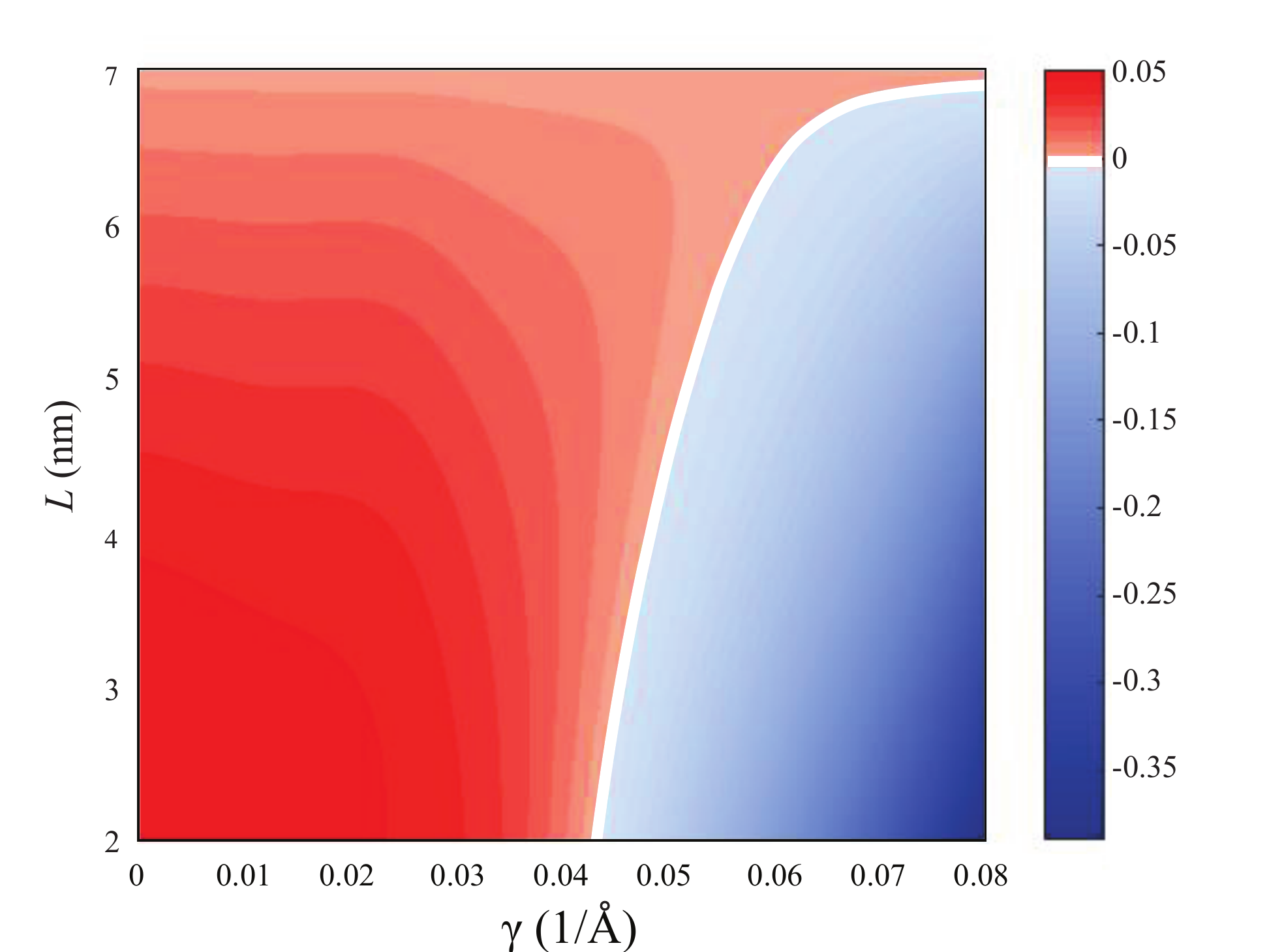}
\caption{\label{phase}(Color online) Phase diagram representing how the value of the gap at the $\Gamma$ point (marked with colors) in the spectrum of surface states of TI depends on parameters of the field. For a given layer thickness one can apparently detect the tendency towards light-induced band inversion (blue area in the right-hand side) upon increasing field intensity and passing through gapless region denoted with a white solid line.}
\end{center}
\end{figure}

For experimentally accessible field intensities the application of an external driving allows us to reduce substantially the value of a gap. The corresponding phase diagram revealing a delicate interplay between the thickness of a thin-film three-dimensional TI and parameters of the field on the value of the gap is plotted in Fig.~\ref{phase}. An interesting observation is that the coupling of the system to an external electromagnetic radiation not only leads to the suppression of anomalous finite-size effects, but also may potentially be exploited to induce band inversion in the spectrum of surface states for rather moderate fields as demonstrated in Fig.~\ref{phase}. In particular, for a given layer thickness, slightly below 7 nm, the system undergoes topological phase transition along the white line in Fig.~\ref{phase} where the gap collapses from a topologically nontrivial phase (red area) to a topologically trivial state (blue area).

\section{Conclusions}\label{sec:conclusions}

In summary, we have investigated anomalous finite-size effects in three-dimensional TIs subjected to an intense off-resonant linearly polarized electromagnetic radiation. The high-frequency expansion reveals that the gap in the spectrum of surface states is quite sensitive to the parameters of the field, and can be suppressed with an increase of intensity within an experimentally accessible range. Intensity of field also affects the components of the Fermi velocity and can lead to the band inversion. These effects may find interesting application in optically controlled spintronics. 

\section*{Acknowledgments}
The support from Megagrant  No. 14.Y26.31.0015 and Project No. 3.8884.2017/8.9 of the Ministry of Education and Science of the Russian Federation is gratefully acknowledged. D.Y. acknowledges the support from RFBR Project No. 16-32-60040. I.A.S. acknowledges support from Project No. 3.2614.2017/4.6 of the Ministry of Education and Science of Russian Federation, Rannis Project No. 163082-051, and Horizon2020 RISE project CoExAN.

\appendix

\section{High-frequency expansion}\label{appendix:expansion}

The Hamiltonian of a three-dimensional topological insulator at the $\Gamma$ point (in the following we focus on Bi$_2$Se$_3$) can be written as

\begin{equation}
H=\mathcal{E}_\mathbf{k}+\begin{pmatrix}
\mathcal{M}_\mathbf{k} & A_1k_z & 0 & A_2k_{-} \\
A_1k_z & -\mathcal{M}_\mathbf{k} & A_2k_- & 0 \\
0 & A_2k_+ & \mathcal{M}_\mathbf{k} & -A_1k_z \\
A_2k_+ & 0 & -A_1k_z & -\mathcal{M}_\mathbf{k} \\
\end{pmatrix},
\end{equation}
where we have defined $\mathcal{E}_k=C+D_1k^2_z+D_2k^2_{\perp}$, $\mathcal{M}_k=B_0-B_1k^2_z-B_2k^2_{\perp}$, $k_\perp^2=k^2_x+k^2_y$, and $k_{\pm}=k_x\pm ik_y$. Being placed to the external field the Hamiltonian acquires time dependence via Peierls substitution $k_y\rightarrow k_y-eE_0\cos(\omega t)/(\hbar\omega)$ (as pointed out in the text we consider a linearly polarized pump field along the $y$ direction). Thus,

\begin{equation}\label{floquet}
H(t)=h_0+2h_1\cos(\omega t)+2h_2\cos(2\omega t),
\end{equation}
where
\begin{equation}
h_0=\mathcal{E}_\mathbf{k}'+\begin{pmatrix}
\mathcal{M}'_\mathbf{k} & A_1k_z & 0 & A_2k_{-} \\
A_1k_z & -\mathcal{M}'_\mathbf{k} & A_2k'_-&0\\
0 & A_2k'_+ & \mathcal{M}'_\mathbf{k} & -A_1k_z \\
A_2k'_+ & 0 & -A_1k_z & -\mathcal{M}'_\mathbf{k} \\
\end{pmatrix},
\end{equation}

\begin{equation}
h_1=-\gamma D_2k_y+\frac{\gamma}{2}\begin{pmatrix}
2B_2k_y & 0 & 0 & iA_2 \\
0 & -2B_2k_y & iA_2 & 0 \\
0 & -iA_2 & 2B_2k_y & 0 \\
-iA_2 & 0 & 0 & -2B_2k_y \\
\end{pmatrix},
\end{equation}

\begin{equation}
h_2=\dfrac{\gamma^2}{4}\begin{pmatrix}
D_2-B_2 & 0 & 0 & 0\\
0 & D_2+B_2 & 0 & 0\\
0 & 0 & D_2-B_2 & 0\\
0 & 0 & 0 & D_2+B_2\\
\end{pmatrix},
\end{equation}
provided that $\mathcal{E}_\mathbf{k}'=C+D_2\gamma^2/2+D_1k^2_z+D_2k_\perp^2$, $\mathcal{M}'_k=B_0-B_2\gamma^2/2-B_1k^2_z-B_2k_\perp^2$. Within the paradigm of high-frequency expansion in the form of Brillouin-Wigner perturbation theory the effective Hamiltonian up to terms proportional to $1/\omega^2$ can be written as follows: 
\begin{equation}
H_\mathrm{eff}=h_0+\dfrac{2}{(\hbar\omega)^2}\sum_{n=1,2}\frac{h_n[h_0,h_n]}{n^2}+\dfrac{[[h_2,h_1],h_1]}{(\hbar\omega)^2}.
\end{equation}
After some algebra we derive the expression (\ref{renham}) of the main text. Note that upon doing high-frequency expansion we project out the Floquet Hamiltonian (\ref{floquet}) to the zero-photon subspace (see Ref. \cite{Mikami2016} for more details). To give a simple quantitative estimation of validity of theoretical formalism developed here we evaluate the maximum instantaneous energy of the time-dependent Hamiltonian (\ref{floquet}) averaged over a period of the field $\frac{1}{T}\int_0^T \mathrm{max}\{\mathrm{spec}||H(t)||\}<\hbar\omega$. Thus, in the vicinity of the $\Gamma$ point the field parameters have to meet  the condition $\gamma\sqrt{A_2^2+B_2^2\gamma^2}/(2\hbar\omega)<1$. Particularly, in the high-frequency regime for an external pumping $\hbar\omega\gtrsim0.25$ eV one can estimate $eE_0\sim0.2(\hbar\omega)^{3/2}$ (eV\,\AA$^{-1}$). While for experimentally accessible parameters, discussed in the main text, $eE_0=0.1$ eV\,\AA$^{-1}$ and $\hbar\omega=1$ eV, one obtains $\gamma=0.1$ \AA$^{-1}$ which quantitatively validates the use of high-frequency expansion.

\section{Derivation of the dispersion relation}\label{appendix:dispersion}

To get the dispersion relation of surface states of a thin slab of TI (infinite in both $x$ and $y$ directions) one must solve the corresponding Schr\"odinger equation $H_\mathrm{eff}(k_x,k_y,-i\partial_z)\psi(z)=E\psi(z)$. The Eigenvalue problem with the trial solution in the form $\psi(z)\sim e^{\lambda z}$ results in

\begin{equation}
\psi_{\pm\alpha1}(z)=\left(\begin{array}{c}
\lambda D_+-L_-+E \\ -i\lambda\tilde{A}_1 \\ 0 \\ \tilde{A}_{2x}k_x+i\tilde{A}_{2y}k_y
\end{array}\right)e^{\pm\lambda_\alpha z}
\end{equation}
and
\begin{equation}
\psi_{\pm\alpha2}(z)=\left(\begin{array}{c}
\tilde{A}_{2x}k_x-i\tilde{A}_{2y}k_y \\ 0 \\ i\lambda\tilde{A}_1 \\ \lambda D_--L_++E
\end{array}\right)e^{\pm\lambda_\alpha z}
\end{equation}
where $D_\pm=D_1\pm B_1$ and $L_\pm=C\pm B_0+(D_2\mp B_2)k_\perp^2$, while $\lambda_\alpha$ ($\alpha=1,2$) are two independent roots of the secular equation $\mathrm{det}||H_\mathrm{eff}(k_x,k_y,-i\lambda)||=0$. Switching to the basis which links the states ($k_x,k_y$) with those of ($-k_x,-k_y$) by inversion determined by
\begin{equation}
\tilde{\psi}_{\pm\alpha\beta}=\frac{1}{2}\left(\psi_{\pm\alpha\beta}\pm(-1)^\beta\psi_{\mp\alpha\beta}\right),
\end{equation}
we search for the general solution in the form
\begin{equation}
\psi(z)=\sum\limits_{\gamma=\pm}\sum\limits_{\alpha,\beta=1,2}C_{\gamma\alpha\beta}\tilde{\psi}_{\gamma\alpha\beta}(z),
\end{equation}
which leads to the dispersion (\ref{disp1}) upon imposing the boundary conditions $\psi(0)=\psi(L)=0$

To solve Eq. (\ref{disp1}) of the main text numerically we look at the points of $(k_x,k_y)$ where (\ref{disp1}) changes sign as a function of $E$ on the condition that the basis functions are linearly independent at this points. The thus obtained solution is shown in Fig.~\ref{dispersion}.

\end{document}